\documentclass[%
 reprint,
 amsmath,amssymb,
 aps, prx,
 longbibliography,
 superscriptaddress,
 altaffilsymbol
]{revtex4-1}
\usepackage[left=2cm, right=2cm, top=3cm, bottom=3cm]{geometry}
\usepackage{amsmath,amssymb}
\usepackage{amsopn}
\usepackage{amsfonts}
\usepackage{natbib,graphicx,url,times,layouts,xcolor}
\usepackage{bm}
\usepackage{dsfont}
\usepackage{bbold}
\usepackage{txfonts} % For PRL fonts
\usepackage{cases} % For numbering in cases format
\usepackage{placeins} %FloatBarrier
\usepackage{textcomp} %mu and other symbols
\usepackage{lineno}
\usepackage{ulem}
\definecolor{blueviolet}{rgb}{0.2, 0.2, 0.6}
\usepackage[pdftex, % FOR PDFLATEX ONLY
bookmarks=false, % Bookmark bar
colorlinks=true,
allcolors=blueviolet,
pdfstartview={FitH}, % FitBH
]{hyperref}

\newcommand{\ket}[1]{{\left\vert{#1}\right\rangle}}

\begin{document}
\title{Multimode storage of quantum microwave fields in electron spins over $100$\,ms}

\author{V.~Ranjan}
\affiliation{Universit\'e Paris-Saclay, CEA, CNRS, SPEC, 91191 Gif-sur-Yvette Cedex, France}

\author{J.~O'Sullivan}
\affiliation{London Centre for Nanotechnology, University College London, London WC1H 0AH, United Kingdom}

\author{E.~Albertinale}
\affiliation{Universit\'e Paris-Saclay, CEA, CNRS, SPEC, 91191 Gif-sur-Yvette Cedex, France}

\author{B.~Albanese}
\affiliation{Universit\'e Paris-Saclay, CEA, CNRS, SPEC, 91191 Gif-sur-Yvette Cedex, France}

\author{T. Chaneli{\`e}re}
\affiliation {Univ. Grenoble Alpes, CNRS, Grenoble INP, Institut N\'eel, 38000 Grenoble, France}

\author{T.~Schenkel}
\affiliation{Accelerator Technology and Applied Physics Division, Lawrence Berkeley National Laboratory, Berkeley, California 94720, USA}

\author{D.~Vion}
\affiliation{Universit\'e Paris-Saclay, CEA, CNRS, SPEC, 91191 Gif-sur-Yvette Cedex, France}

\author{D.~Esteve}
\affiliation{Universit\'e Paris-Saclay, CEA, CNRS, SPEC, 91191 Gif-sur-Yvette Cedex, France}

\author{E.~Flurin}
\affiliation{Universit\'e Paris-Saclay, CEA, CNRS, SPEC, 91191 Gif-sur-Yvette Cedex, France}

\author{J.~J.~L.~Morton}
\affiliation{London Centre for Nanotechnology, University College London, London WC1H 0AH, United Kingdom}

\author{P.~Bertet}
\email{patrice.bertet@cea.fr}
\affiliation{Universit\'e Paris-Saclay, CEA, CNRS, SPEC, 91191 Gif-sur-Yvette Cedex, France}

\newcommand{\Tphon}{T_\text{phon}}
\newcommand{\Tphot}{T_\text{phot}}
\newcommand{\Tphotc}{T_\text{phot}^\text{cold}}
\newcommand{\Tphoth}{T_\text{phot}^\text{hot}}
\newcommand{\Ti}{T_\text{int}}
\newcommand{\Tih}{T_\text{int}^\text{hot}}
\newcommand{\Tic}{T_\text{int}^\text{cold}}

\newcommand{\ki}{\kappa_\text{int}}
\newcommand{\ke}{\kappa_\text{ext}}

\newcommand{\gens}{g_\text{ens}}
\newcommand{\Nspins}{N_\text{spins}}
\newcommand{\TE}{T_2^\text{E}}
\newcommand{\TIE}{T_1^\text{E}}
\newcommand{\dfdb}{\text{d}f/\text{d}B_0 }

\maketitle

\textbf{A long-lived multi-mode qubit register is an enabling technology for modular quantum computing architectures. For interfacing with superconducting qubits, such a quantum memory should be able to store incoming quantum microwave fields at the single-photon level for long periods of time, and retrieve them on-demand. Here, we demonstrate the partial absorption of a train of weak microwave fields in an ensemble of bismuth donor spins in silicon, their storage for 100ms, and their retrieval, using a Hahn-echo-like protocol. The long storage time is obtained by biasing the bismuth donors at a clock transition. Phase coherence and quantum statistics are preserved in the storage.}

%%introduction#############################
Quantum memory as a matter-based information storage medium for itinerant qubits has been recognised a powerful ingredient in quantum technologies, underpinning applications such as quantum repeaters~\cite{lvovsky_optical_2009}. In analogy with memories in classical computing, quantum memories offer storage that is both long-term compared to data lifetimes in processing qubits as well as high density, for example when multi-mode memories are employed to store a large number of states. These attributes can be of general benefit to quantum computing architectures, supporting approaches with a high degree of modularity. Inspired by such possibilities, quantum memories in the optical domain have been developed in particular using rare-earth-ion-doped crystals, reaching high efficiency~\cite{hedges_efficient_2010}, and storage times in the millisecond range~\cite{businger_optical_2020}.

\begin{figure}[t!]
  \includegraphics[width=\columnwidth]{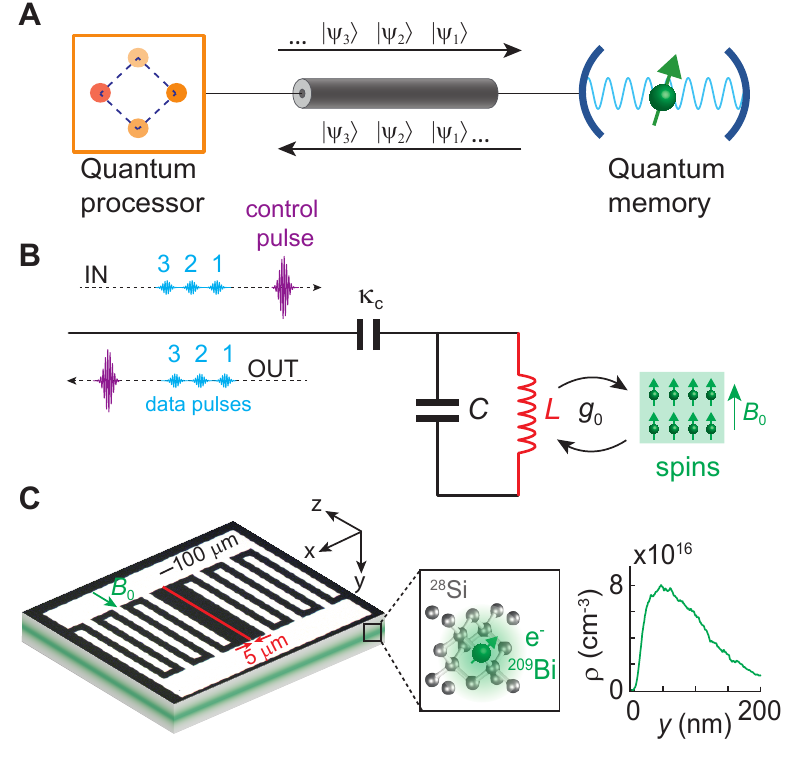}
  \caption{\label{fig:Exptsetup}
  \textbf{Quantum memory. (A)} The proposed architecture consisting of a  quantum processor coupled via a coaxial cable to a quantum memory from electron spins. \textbf{(B)} Circuit implementation of the quantum memory:  an ensemble of $N$ electron spins are inductively coupled to a lumped resonator which is capacitively coupled  to a microwave line. Weak coherent data pulses in our demonstration, travel along the lines, are stored by the spin ensemble, and then released on-demand by a control pulse. The coupling strength of an individual spin to the resonator is $g_0$. \textbf{(C)} The hybrid resonator-spin system. The resonator is fabricated on top of the silicon substrate in a superconducting aluminum thin film to minimize internal losses. It consists of a capacitor shunted by a $5~\mu \mathrm{m}$-wide inductance wire, to which the spin of bismuth donors implanted around a depth $\sim 100~$nm are inductively coupled. A magnetic field $B_0$ is applied along the inductor ($z$ direction).}
\end{figure}

Quantum memories suitable for interfacing with superconducting quantum processors, must instead operate in the \textit{microwave} regime, which requires operation at millikelvin temperatures in a dilution refrigerator. A microwave multimode quantum memory with long storage times would represent a potent and versatile new component in quantum computing architectures based on superconducting qubits. For example, it could be used to realise sub-processors operating a Quantum Turing Machine architecture with high internal connectivity and built-in long-term memory [see Fig.\ref{fig:Exptsetup}\textbf{(A)}]~\cite{tordrup_holographic_2008}, helping to overcome some of the limitations of present day superconducting qubit processors~\cite{rigetti_superconducting_2012,barends_coherent_2013,arute_quantum_2019}.

%Present-day superconducting quantum processor prototypes based on arrays of transmon qubits suffer from short qubit coherence time, limited connectivity, and from the need to have as many control and readout lines as qubits. A long-lived qubit register, able to store and retrieve on-demand a large number of qubit states with negligible decoherence, could help overcoming these short-comings. 

For implementing such a quantum memory, superconducting microwave cavities~\cite{reagor_quantum_2016,naik_random_2017} and mechanical resonators~\cite{palomaki_coherent_2013,hann_hardware-efficient_2019} have been considered, with storage times in the millisecond range. Ensembles of electron spins in solids offer a large number of degrees of freedom well decoupled from their environment with coherence times that can reach seconds~\cite{tyryshkin_electron_2012,steger_quantum_2012,muhonen_storing_2014}, and are thus well suited to implement a many-mode quantum memory with long storage time~\cite{tordrup_holographic_2008,wesenberg_quantum_2009}. For modularity, it is natural to physically separate the quantum processor and the quantum memory, and to interface the two devices via propagating microwave photons. Operating a spin-ensemble-based quantum memory thus amounts to absorbing incoming microwave photons and releasing them on-demand in the same quantum state [see Fig.~\ref{fig:Exptsetup}(\textbf{A})]. 

A convenient way to interface the spins and the incoming photons [see Fig.~\ref{fig:Exptsetup}\textbf{(B)}] is via a superconducting micro-resonator of frequency $\omega_0$, capacitively coupled to the input line with an energy damping rate $\kappa$, and inductively coupled (with single spin coupling strength $g_0$) to an ensemble of $N$ spins, characterized by its Larmor frequency $\omega_S$ with inhomogeneous linewidth of Full-Width-Half-Maximum $\Gamma$. The resonator serves to enhance microwave absorption and re-emission by spins, but also provides a convenient reset mechanism for the memory, via the Purcell relaxation of each individual spin at a rate $\Gamma_P = 4 g_0^2 / \kappa$~\cite{purcell_spontaneous_1946,bienfait_controlling_2016}.

The physics of the memory can be demonstrated using weak resonant coherent pulses with a small average photon number. Such microwave pulses with amplitude envelope $\beta_\text{in}(t)$ are absorbed by the spins with an efficiency governed by the ensemble cooperativity $C =  N \Gamma_P /  \Gamma$. Since the reflected pulse amplitude is $\beta_\text{ref}(t) = \beta_\text{in}(t) (1 - C)/(1+C)$, complete absorption is achieved for $C=1$, which appears as a necessary condition for a high-fidelity memory (see \cite{afzelius_proposal_2013} and Supplementary Materials). After absorption in the spin ensemble, the microwave fields should be retrieved using sequences of control pulses. The simplest sequence consists in applying a $\pi$ pulse to the spins after a delay $\tau$, which generates an echo of the absorbed pulse at time $2\tau$. Because this echo is generated at a time when nearly all spins are in the excited state, it is unavoidably accompanied by $N \Gamma_P / \Gamma = C$ spontaneously emitted noise photons, thus reducing the memory fidelity~\cite{ruggiero_why_2009}. Therefore a more complex protocol must be used, involving two $\pi$ pulses and dynamic control of the cavity frequency, in order to form the echo in the spin ground state and thus avoid added noise~\cite{afzelius_proposal_2013}.

%Because these echoes are generated while the spins are nearly all in the excited state, spontaneous emission may however add extra noise to the echo, thus reducing the memory fidelity. In the limit $C \ll 1$, this added noise is negligible, and the emitted echo has the statistics of a coherent state, as the input field. When $C = 1$ however, this is no longer the case, and a more complex protocol must be used, involving two $\pi$ pulses and dynamic control of the cavity frequency, in order to form the echo in the spin ground state without added noise~\cite{afzelius_proposal_2013}. %

%To retrieve the stored excitation after a time $2 \tau$, a microwave $\pi$ pulse is applied to the spins at time $\tau$, resulting in the emission of an echo. The efficiency of the whole protocol is governed by the ensemble cooperativity $C = 4 g_0^2 N / (\kappa \Gamma)$; indeed, the amplitude of the reflected pulse and of the echo is $\beta_{ref}(t) = \beta_{in}(t) (1 - C)/(1+C)$ and $\beta_{echo} = \beta_{in}(t) 4 C / (1+C)^2$ respectively (assuming an input pulse bandwidth smaller than $\kappa$ and $\Gamma$\cite{afzelius_proposal_2013}, and a resonator without internal losses). 

Reaching unit cooperativity requires large spin concentrations; but spin-spin interactions then reduce the coherence time. This proved to be a serious limitation in previous experiments storing microwaves in spin ensembles, first in the classical regime~\cite{wu_storage_2010}, then in the quantum regime~\cite{grezes_multimode_2014,probst_microwave_2015,grezes_storage_2015}, where the longest storage times demonstrated reached only of order $100~\mu\mathrm{s}$. This conflict can be mitigated by biasing the spins at specific magnetic fields where their effective magnetic moment vanishes (thus minimizing decoherence induced by spin-spin interactions), while keeping a finite transverse susceptibility so that $g_0$ remains non-zero, opening the possibility to reach $C=1$ without compromising the coherence time. Such ``clock transition" (CT) or ``ZEro-First-Order-Zeeman" (ZEFOZ) points occur in spin systems where the electron spin is strongly hybridized with a nuclear spin by the hyperfine interaction, as in bismuth donors in silicon~\cite{mohammady_bismuth_2010,wolfowicz_atomic_2013} and rare-earth-ion-doped crystals~\cite{zhong_optically_2015,ortu_simultaneous_2018}. 

%The spin-resonator coupling strength is $g_0 = \gamma_e \delta B_{1} \langle 0 | S_x |1 \rangle$, $\delta B_1$ being the rms vacuum fluctuations of the resonator $B_1$ field.

\begin{figure}[t!]
  \includegraphics[width=\columnwidth]{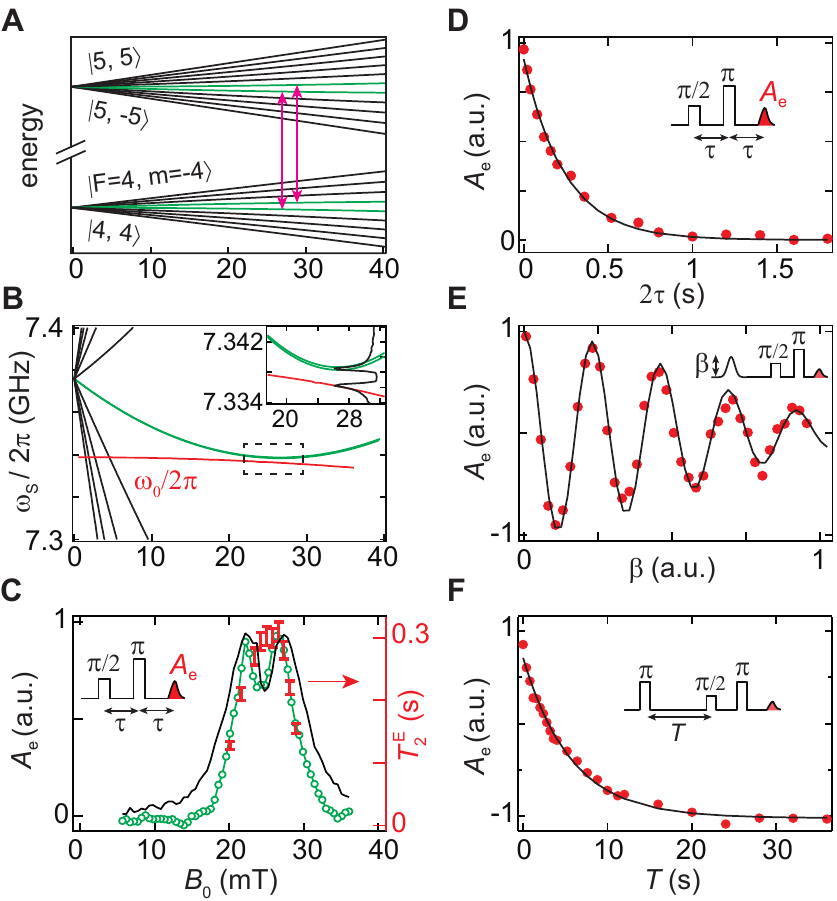}
  \caption{ \label{fig:bismuth}
  \textbf{Clock transition in bismuth and its characterization. (A)} Energy level diagram $|F,m\rangle$ of electron spins from bismuth at small magnetic fields. Green lines indicate the four levels $|4,-1\rangle,~ |4,0\rangle,~ |5,-1\rangle$, and $|5,0\rangle$ involved in the two quasi-degenerate CT. \textbf{(B)} The ESR-allowed transition frequency of bismuth donor spins in silicon between $7.3-7.4$~GHz, and the measured resonator frequency $\omega_0$ as a function of $B_0$. The CT with $\text{d}\omega/\text{d}B_0 =0$ can be seen at $B_0 = 27~$mT. Inset: zoom around CT to show degenerate transition and expected strain broadened spectrum in frequency. \textbf{(C)} Measured echo detected spectroscopy around the CT (green symbols, left axis), calculated one (black line, left axis - see Supplementary), and two-pulse echo coherence time $T_2^E$ versus $B_0$ (red bars, right axis). \textbf{(D)}  Measured echo area decay at CT (symbols) and corresponding exponential fit (line) yielding $T_2^E = 0.3 \pm 0.05$~s. Magnitude detection is employed to circumvent phase noise from the measurement setup. \textbf{(E)} Measured (symbols) and simulated (line) Rabi oscillations at the CT. Each $A_e$ is plotted as a function of the amplitude $\beta$ of the first pulse. \textbf{(F)} Measured spin relaxation at the CT using an inversion recovery sequence (symbols), and corresponding exponential fit (line) yielding $\TIE = 5.5 \pm 0.5$~s. }
\end{figure}

%% Bismuth+ Resonator #####################
Here, we use an ensemble of bismuth donors in silicon biased at a clock transition to demonstrate the long-term storage of microwave fields. The device schematic is shown in Fig.~\ref{fig:Exptsetup}\textbf{(C)}. Bismuth atoms were implanted around a $\sim 100$\,nm depth in a silicon substrate that was enriched with the nuclear-spin-free $^{28}$Si isotope for longer coherence time. At low temperature, bismuth atoms trap a conduction electron, forming the donor systems. The spin Hamiltonian $H_\text{Bi}/\hbar = (\gamma_\text{e} \bold{S} + \gamma_\text{n} \bold{I})\cdot \bold{B_0} + A \bold{S} \cdot \bold{I}$ is the sum of the Zeeman interaction of the electron (nuclear) spin $S=1/2$ ($I=9/2$) of the bismuth donor with the applied magnetic field $B_0 $ ($\gamma_\text{e}/2\pi \simeq 28$~MHz/mT and $\gamma_\text{n}/2\pi \simeq 7$~kHz/mT being the electronic and nuclear gyromagnetic ratios) and of their hyperfine interaction with a strength $A/2\pi = 1.475~$GHz. The resulting energy levels $\ket{F,m}$ can be grouped in a low-energy ($F=4$) manifold of $9$ states and a high-energy ($F=5$) manifold of $11$ states [see Fig.~\ref{fig:bismuth}\textbf{(A)}], separated by $\sim 7.38$\,GHz, $m$ being the eigenvalue of the total angular momentum $S_z + I_z$ along the field direction $z$~\cite{mohammady_bismuth_2010}. The operator $S_x$ has non-zero matrix elements between all pairs of states that verify $\Delta m = \pm 1$, and transitions between such states are therefore allowed under a transverse driving microwave field $B_1$ along the $x$ direction. We note that transitions $\ket{4,m} \leftrightarrow \ket{5,m-1}$ and $\ket{4,m-1} \leftrightarrow \ket{5,m}$ are quasi-degenerate. In this work we are interested in the $\ket{4,0} \leftrightarrow \ket{5,-1}$ and $\ket{4,-1} \leftrightarrow \ket{5,0}$ transitions, which satisfy the CT condition at $\sim 7.338$\,GHz and $B_0 = 27$\,mT where $d\omega/dB_0 = 0$ while the transition matrix elements $\langle 4,0 |S_x | 5,-1\rangle =\langle 4,-1 |S_x | 5,0\rangle= 0.25$ remain non-zero. To describe the interaction with microwave fields close to resonance, we model the pair of transitions as independent spin-1/2 systems labelled generically as $\ket{0}$ ($\ket{1}$) for the ground (excited) state.

%The resonator is fabricated directly on top of the silicon substrate containing the spins, in a superconducting aluminum thin film to minimize internal losses. It consists in a capacitor shunted by an inductance wire of $5 \mu \mathrm{m}$ width, to which the bismuth donors are inductively coupled. A magnetic field $B_0$ is applied parallel to the sample, along the inductor ($z$ direction). The spin-resonator coupling strength is $g_0 = \gamma_e \delta B_{1} \langle 0 | S_x |1 \rangle$, $\delta B_1$ being the rms vacuum fluctuations of the $B_1$ field.

The resonator is designed such that its resonance $\omega_0$ is close to the CT frequency of $7.338$\,GHz. A finer tuning of $\omega_0$ is obtained by changing the resonator coupling to the measurement line, which is controlled by the length of a microwave antenna inserted in the sample holder containing the silicon chip (see Supplementary Materials). Figure \ref{fig:bismuth}\textbf{(B)} shows $\omega_0$ as a function of $B_0$. Due to the kinetic inductance contribution to the resonator inductance, $\omega_0/2\pi$ decreases with $B_0$, reaching $7.336$\,GHz at $27$\,mT. The external energy coupling rate of the resonator is $\kappa_c = 4 \times 10^5~\text{s}^{-1}$, though its total damping rate, $\kappa = \kappa_c + \kappa_i$, including internal losses is power dependent (see Supplementary Materials). We find that at low input powers corresponding to one intra-cavity photon on average, $\kappa_c/\kappa \sim 0.3$, whereas at high power $\kappa_c/\kappa \sim 0.5$, revealing that part of the losses are caused by Two-Level-Systems (TLS). The experiments are performed at $T=20$\,mK, in the quantum regime for microwave fields $\hbar \omega_0 \gg k_\text{B} T$.

Spin spectroscopy is performed with a custom-built spectrometer described in more details in Ref.~\onlinecite{bienfait_reaching_2016}, using the micro-resonator for the inductive spin detection. Hahn-echo sequences of pulses ($\pi/2 - \tau - \pi - \tau - echo$) are sent to the resonator input at $\omega_0$. The resulting echo is amplified by a Josephson travelling wave parametric amplifier (JTWPA) at $20$\,mK that adds noise close to the quantum limit~\cite{macklin_nearquantum-limited_2015}, before further amplification at higher temperatures and demodulation at room-temperature. Figure~ \ref{fig:bismuth}\textbf{(C)} shows that an echo signal is observed over a range of $B_0$ values around the CT, despite the detuning between the resonator frequency and the expected donor frequency. This is explained by the differential thermal contractions of the resonator thin film with respect to the underlying substrate, which causes spatial variations of the strain profile and consequently of the donor hyperfine constant and the Larmor frequency~\cite{pla_strain-induced_2018}. Based on the measured lineshape for the first transition near 1.5~mT [see inset of Fig.~\ref{fig:bismuth}\textbf{(B)}], the echo-signal field dependence of Fig.~\ref{fig:bismuth}\textbf{(C)} is semi-quantitatively accounted for (see Supplementary Materials for more details on lineshape). 
%Interestingly, strain broadening also relaxes the need for precise control of the resonator frequency to reach the CT frequency.

\begin{figure*}[t!]
  \includegraphics[width=2\columnwidth]{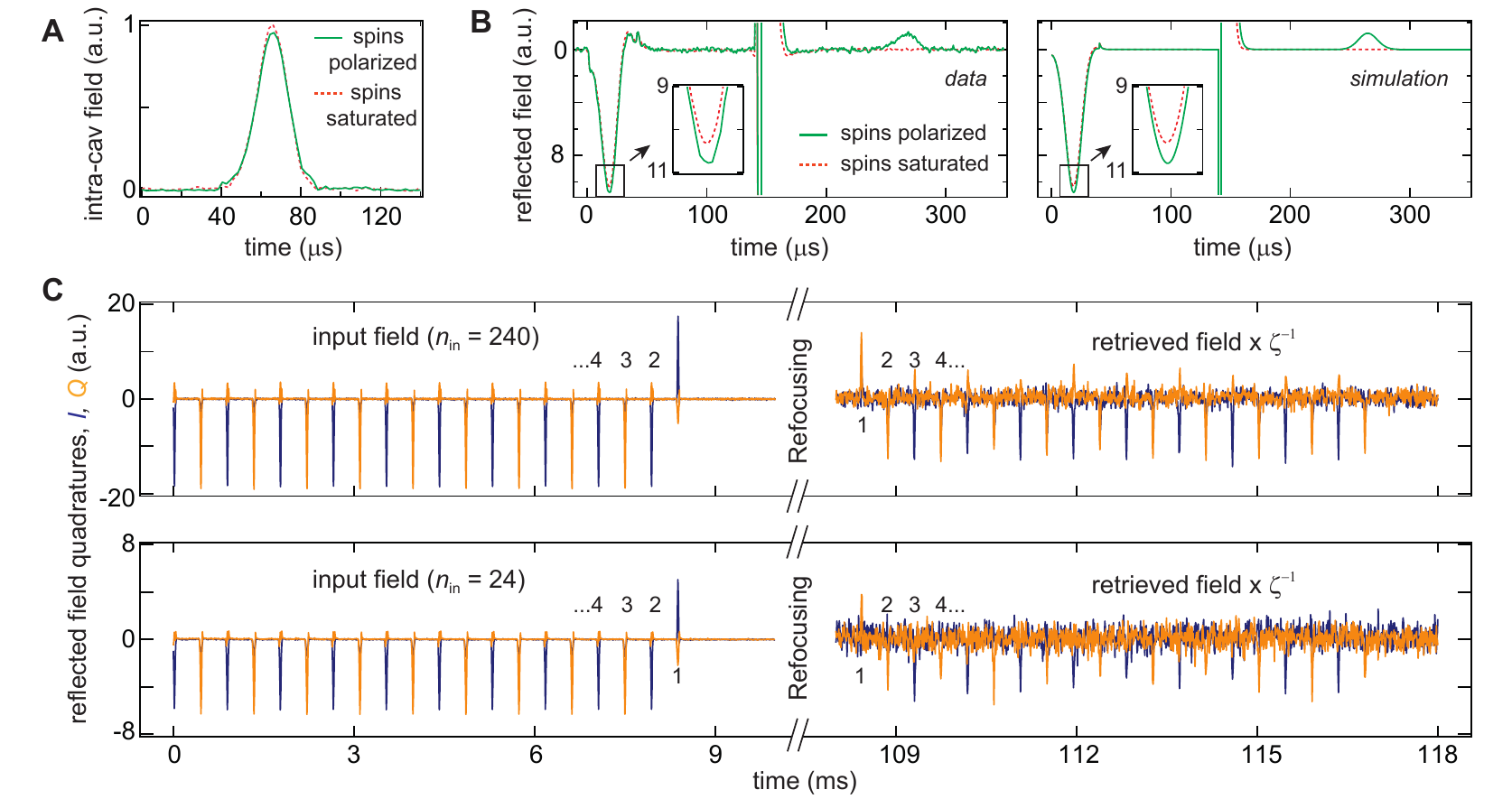}
  \caption{\label{fig:memory}
  \textbf{Storage and retrieval of weak pulses at clock transition}. \textbf{(A)} Measured intra-cavity field $\alpha$ averaged over $10^3$ repetitions at a repetition rate of 200~ms~(spins saturated, dashed curve) and of 16~s (spins polarized, solid curve).  \textbf{(B)} Measured (left panel, $4\times 10^3$ averages) and simulated (right panel) reflected field on resonance ($\sqrt{\kappa_c}\alpha-\beta_\text{in}$) and echo at a time $\ll T_2^\text{E}$. The simulation assumes a spectral density of $N/(\Gamma/2\pi) = 10^6~\text{spins}/$MHz and a fixed $g_0/2\pi=40$~Hz. For panels (A, B) there are 240 photons in the input pulse. \textbf{(C)} Top (bottom): a train of weak microwave fields, with input field $\beta_\text{in}$ measured off resonantly to allow comparison,
  containing 240 (24) photons at the input and their retrieval after the refocusing $\pi$ pulse. The numbering highlights that retrieval maintains the phase relation with respect to the input field. The retrieved fields are multiplied by $\zeta^{-1}\approx 24$, where $\zeta= 4C(\kappa_C/\kappa)$ is the theoretical efficiency of the retrieved field at a time $\ll T_2^E$. 
  }
\end{figure*}

%% Clock transition characterization #####################
We then measure the spin coherence time at the CT by recording the integral of an echo as a function of the delay $2 \tau$ [see Fig.~\ref{fig:bismuth}\textbf{(D)}], which is well fitted by an exponential yielding $\TE = 0.30$~s. This is the longest coherence time measured for electron spins in a nanostructure, in agreement with the $2.7$\,s measured in the bulk at another bismuth CT and a lower bismuth concentration~\cite{wolfowicz_atomic_2013}. As expected, $\TE$ is longest near $B_0 = 27$\,mT where $\text{d}\omega/ \text{d}B_0=0$ [see Fig.~\ref{fig:bismuth}\textbf{(C)}, right axis]; also, for comparison, we measure $\TE = 7.5$\,ms on the $\ket{4,-4}\leftrightarrow\ket{5,-5}$ transition at $B_0 = 1.4$\,mT with $\text{d}\omega/ \text{d}B_0 \sim \gamma_e$, which confirms the interest of CTs for long coherence times (a detailed study of the coherence time will be presented elsewhere). The Purcell spin relaxation time $T_1 = 5.5$\,s is measured using an inversion recovery sequence [see Fig.~\ref{fig:bismuth}\textbf{(F)}], yielding a spin-photon coupling constant $g_0 /2\pi = 40$\,Hz. All the experiments described in the following are therefore averaged with a repetition time of $3T_1 \sim 16$\,s.

We now demonstrate the coherent absorption and retrieval of weak microwave pulses by the donors at the CT. A Gaussian incoming pulse envelope $\beta_\text{in}(t)=\beta \exp [-(t/t_0)^2]$ is chosen, with $t_0 = 10~\mu \mathrm{s}$ larger than the cavity field decay time $2\kappa^{-1} = 1.5~\mu \mathrm{s}$. To calibrate the input pulse intensity in photon number $n_\text{in}=\int |\beta_\text{in}(t)|^2 dt$, we compare the measured Rabi nutation of the donor spins as a function of $\beta$ [see Fig.~\ref{fig:bismuth}\textbf{(E)}], with numerical simulations in which all the parameters are determined experimentally. Next, we determine the ensemble cooperativity by measuring the ratio $\alpha^p(t) / \alpha^s(t) = 1/(1+C)$ between the intra-cavity field $\alpha^p(t)$ (resp. $\alpha^s(t)$) with spins polarized (resp. unpolarized). The data shown in Fig.~\ref{fig:memory}\textbf{(A)} yield $C = 3.5 \pm 1 \times 10^{-2}$, corresponding to a spin density $N / (\Gamma/2\pi) \sim 10^6~\text{spins}/\mathrm{MHz}$ consistent with the known sample implantation parameters. Note that a careful account of the contribution from resonator losses and TLS was necessary in the analysis (see Supplementary Materials). 

%We note that the ability to perform well-defined rotations as shown in Fig.~\ref{fig:bismuth} is essential for a quantum memory protocol; in our experiment, this is enabled by the strain gradient~\cite{pla_strain-induced_2018}. We then use numerical simulations without adjustable parameters to compute the input pulse photon number $\int |\beta(t)|^2 dt$. 

%The nutation frequency is $2 g_0 \sqrt{\bar{n}}$, $\bar{n}$ being the average intra-resonator photon number. We thus obtain the intra-cavity field photon number $\bar{n}$, from which we deduce the input pulse photon number $n_\text{cav}(\kappa/\Omega_\text{in})$ (CHECK). 

%The absorption of a low-power microwave pulse by the spins is measured by comparing the intra-cavity field $\alpha^p(t)$ measured when the effective two-level spin systems are fully polarised, with that measured $\alpha^s(t)$ when the effective spins are fully saturated and hence not interacting with the field. Typical data are shown in Fig.~\ref{fig:memory}\textbf{(A)}, for an input pulse containing $ n_\text{in} = 240 \pm 24 $  photons. The ratio of the fields $\alpha^p(t)/\alpha^s(t)$ directly yields the cooperativity $C = 3.5 \pm 1 \times 10^{-2}$ and thus the absorption efficiency  $\zeta=4C(\kappa_c/\kappa) = 4.3 \pm 1.4 \times 10^{-2}$ (see Supplementary).%

With the knowledge of the cooperativity $C$, the whole storage/retrieval protocol can be demonstrated and quantitatively understood. Because $C \ll 1$, we use here the simple retrieval protocol based on a single $\pi$ pulse, with a square shape of 2~$\mu$s duration. For $n_\text{in} = 240 \pm 24$ input photons, $\zeta n_\text{in}$ should be absorbed, with $\zeta=4C(\kappa_c/\kappa) = 4.3 \pm 1.4 \times 10^{-2}$ (see Supplementary Materials), and $\zeta^2 n_\text{in} = 0.45 \pm 0.15$ photons should be re-emitted as a Hahn echo. Data shown in Fig.~\ref{fig:memory}\textbf{(B)} for $\tau = 100~\mu$s are in quantitative agreement with these analytical predictions as well as with a complete simulation of the experiment.

%The cooperativity This corresponds to $\sim 10 \pm 3$ photons being absorbed in the spin ensemble. Note that careful account of the contribution from resonator losses and two level systems was necessary in the analysis (see Supplementary). From $C$, we obtain a spin density of $N / (\Gamma/2\pi) \sim 10^6~\text{spins}/\mathrm{MHz}$.%

%Because $C \ll 1$, we use the simple retrieval protocol based on a single $\pi$ pulse (square shape) of 2~$\mu$s duration. Subject to a refocusing pulse at a delay $\ll T_2^E$, reflected field $\beta_\text{ref}(t)$ of an incident pulse containing $n_\text{in} = 240 \pm 24$ photons and the corresponding echo is shown in Fig.~\ref{fig:memory}\textbf{(B)}. For comparison, the case of saturated spins is also shown. These measurements are reliably reproduced with numerical simulations using the experimentally extracted spin density $N /\Gamma$ and the single spin-photon coupling strength $g_0$.

%Because $C \ll 1$, we use the simple retrieval protocol based on a single $\pi$ pulse of 2~\textmu s duration, leading to an echo shown in Fig.~\ref{fig:memory}\textbf{(B)}. As expected in the limit $C \ll 1$, the echo amplitude is nearly twice larger than the absorbed field (see Supplementary). The simulation reproduces the echo amplitude quantitatively (see Fig.\ref{fig:memory}\textbf{(B)}). 

We then demonstrate long-lived and multi-mode first-in/last-out microwave storage by sending a train of $20$ weak Gaussian pulses with varying phases, and retrieving them using a single refocusing pulse. The experiment was performed twice, comparing input pulse intensities of $n_\text{in} = 240 \pm 24$ and $24 \pm 2$ photons. As seen in Fig.~\ref{fig:memory}\textbf{(C)}, echoes are retrieved after 100~ms with a well-defined phase. The retrieved field amplitude is slightly reduced from the expected value of $\zeta \beta_\text{in}$, mostly due to spin decoherence during the storage time, and also for a small part to resonator phase noise caused by vortices trapped in the resonator thin film (see Supplementary Materials). The recovered intensity corresponds to respectively $0.3 \pm 0.1$ and $0.03 \pm 0.01$ photons per echo, $\sim 10^{-3}$ of the input pulse energy.

\begin{figure}[t!]
  \includegraphics[width=\columnwidth]{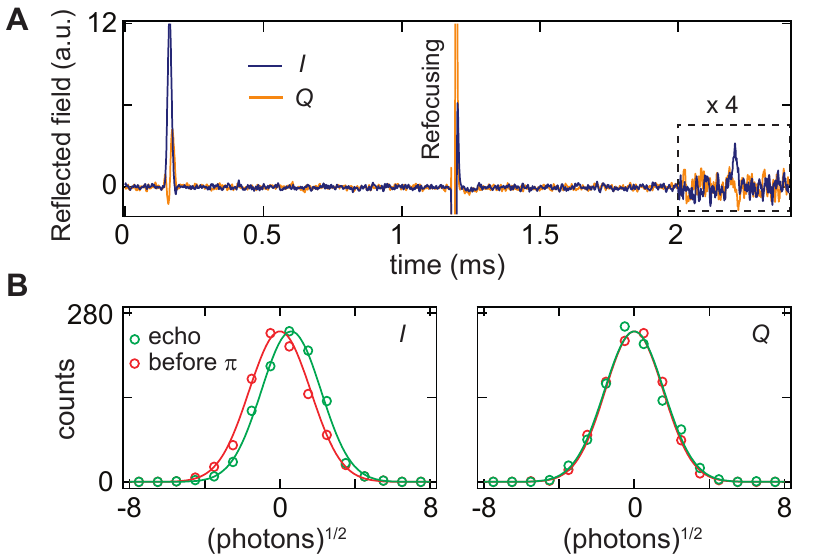}
  \caption{\label{fig:noise}
  \textbf{Noise statistics of retrieved echo  (A)} Average of $10^3$ single shot traces of reflected signals ($\sqrt{\kappa_c}\alpha-\beta_\text{in}$) acquired with a repetition rate of 16~s. The input and retrieved field contain $240$ photons and 0.25~photons respectively. In this run, the internal losses were slightly larger than in Fig.~\ref{fig:memory}\textbf{(B)}, leading to a lower value of $\zeta$ and a lower echo amplitude. \textbf{(B)} Histograms of signal quadratures before the refocusing $\pi$ pulse and on the echo. Each histogram sample is acquired by integrating signals for $100~\mu$s weighted by a normalized gaussian mode shape describing the echo. Solid curves are calculated gaussians of same area.}
\end{figure}

%% ############################ Outlook + Noise in echo + conclusion ######
Finally, we address the question of the quantum statistics of the echo field. For that, we record a histogram of integrated output signals acquired during the echo $E$, and outside the echo $O$ at a time when all spins are in their ground state. Thanks to the photon number calibration, the average echo amplitude $\bar{E} - \bar{O}= \zeta \sqrt{n_\text{in}} = 0.5 \pm 0.1$ provides an absolute calibration of the horizontal scale in dimensionless (square root of photon number) units. This enables a comparison of the measured standard deviations $\sigma_O \sim \sigma_E = 1.5 \pm 0.3$ to the expected $\sigma_O = \sqrt{(\bar{n}_\text{id} + 1)/2}$ and $\sigma_E = \sqrt{(\bar{n}_\text{id} + C + 1)/2}$ (see supplementary Materials), where $\bar{n}_\text{id}$ describes JTWPA non-ideality, microwave losses between the sample and the amplifier, and added noise by the higher-temperature amplification chain. We find $\bar{n}_\text{id} = 3.5 \pm 1.7$, a reasonable value compared to those measured in similar circuit QED setups. Then, no measurable difference is observed between $\sigma_E$ and $\sigma_O$, in agreement with the theoretical estimate which predicts less than $0.5\%$ difference. Overall, these measurements prove the consistency between our photon number calibration protocol and the statistics of the recovered signal, and they show that the echo is recovered with negligible added noise, close to the quantum limit.

%The noise per quadrature with contributions from vacuum fluctuations, spontaneous emission during the echo and quantum-limited TWPA amplifier is expected to be $\sqrt{(2+C)/4}$ (see Supplementary). Since $C \ll 1$, the echo statistics should be close to a coherent state. A histogram of integrated echo signals $E$ for an input pulse with $n_\text{in}= 240$ photons is shown in Fig.~\ref{fig:noise}\textbf{(B)} and compared to the histogram of integrated signals outside of the echo, before the $\pi$ pulse. We note that total $Q$ for this measurement was $20\%$ smaller compared to Fig~3\textbf{(B)} leading to a $40\%$ smaller echo that now contains $\sim 0.25$ photons. No measurable difference is found in standard deviation $\sigma_E$ of histograms, confirming negligible added noise during the echo. The shift of distributions yield signal-to-noise ratio $\bar{E} / \sigma_E = 0.32$. Using the photon number calibration, we deduce that $\bar{E} = \sqrt{0.25}= 0.5 \pm 0.1$, yielding the noise level in absolute units $\sigma_E = 1.5 \pm 0.3$. This is a factor $2.1 \pm 0.7$ above the quantum limit $\sigma_E = 1/\sqrt{2}$, likely due to attenuation and reflections between the sample and the JTWPA, added noise by the JTWPA, and added noise by the higher-temperature amplification chain. Overall, these measurements prove the consistency between our photon number calibration protocol and the statistics of the recovered signal, and they show that the echo is recovered with negligible added noise, close to the quantum limit.

%%################ outlook and conclusion  #################
Turning this proof-of-principle into an operational quantum memory requires increasing the ensemble cooperativity by a factor $\sim 30$, up to $C=1$. We argue that this can be achieved by straightforward design adjustments and without compromising the spin coherence time, by increasing both the resonator quality factor and the number of implanted spins. Importantly, we propose to increase the number of spins at fixed concentration, simply using a deeper implantation profile, up to $1~\mu \mathrm{m}$ as already shown~\cite{albanese_radiative_2020}. The demonstration of a fully operational microwave quantum memory will then be achieved using the two-pulse protocol proposed in~\cite{afzelius_proposal_2013}, and applied to store quantum states originating from a transmon-based quantum processor.

In conclusion, we have demonstrated the absorption of trains of weak microwave pulses consisting of only a few photons in a hybrid quantum device, their storage for $100$\,ms, and their phase-coherent re-emission without added noise. Our results illustrate the utility of clock transitions for efficient quantum memories with long storage times as well as memory reset via the Purcell effect.

%In this high-cooperativity regime however, the simple protocol described here will need to be revisited to avoid added noise during echo emission \cite{afzelius_proposal_2013,julsgaard_quantum_2013}.

%%################ Refernces #################
%\bibliographystyle{unsrt}
%\bibliography{ClockTransitionMemory}

%%############### Acknowledgements ###########
\section*{Acknowledgements}
We thank P. Se\'nat, D. Duet and J.-C. Tack for the technical support, and are grateful for fruitful discussions within the Quantronics group. We acknowledge IARPA and Lincoln Labs for providing a JTWPA used in the measurements. We acknowledge support from the Horizon 2020 research and innovation program through grant agreement No. 771493 (LOQO-MOTIONS), and of the European Union through the Marie 365 Sklodowska Curie Grant Agreement No. 765267 (QuSCO), and of the Agence Nationale de la Recherche (ANR) through projects QIPSE, MIRESPIN (ANR 19 CE47 0011), and the Chaire Industrielle NASNIQ, and of the UK's Engineering and Physical Sciences Research Council through a Doctoral Training Award. T. S. was supported by the U.S. Department of Energy under Contract No. DE-AC02-05CH11231.

%%############## Author contributions ##########
\section*{Author contributions}
V.R, J.J.M and P.B designed the experiment. T.S provided the implanted Si sample on which V.R fabricated the device. V.R and J.S performed  measurements and numerical simulations with inputs from E.A and B.A. T.C provided theory support. V.R and P.B wrote the manuscript with useful contribution from  all co-authors. 

%################### SUPPLEMENTARY ###################
\newpage
\textbf{SUPPLEMENTARY MATERIAL}
\newline

\renewcommand{\theequation}{S\arabic{equation}}
\renewcommand{\thefigure}{S\arabic{figure}}
\renewcommand{\thetable}{S\arabic{table}}

\setcounter{equation}{0}
\setcounter{figure}{0}

%################### Resonator and pulses ###################
\section{Device setup}
\begin{figure}[t!]
  \includegraphics[width=\columnwidth]{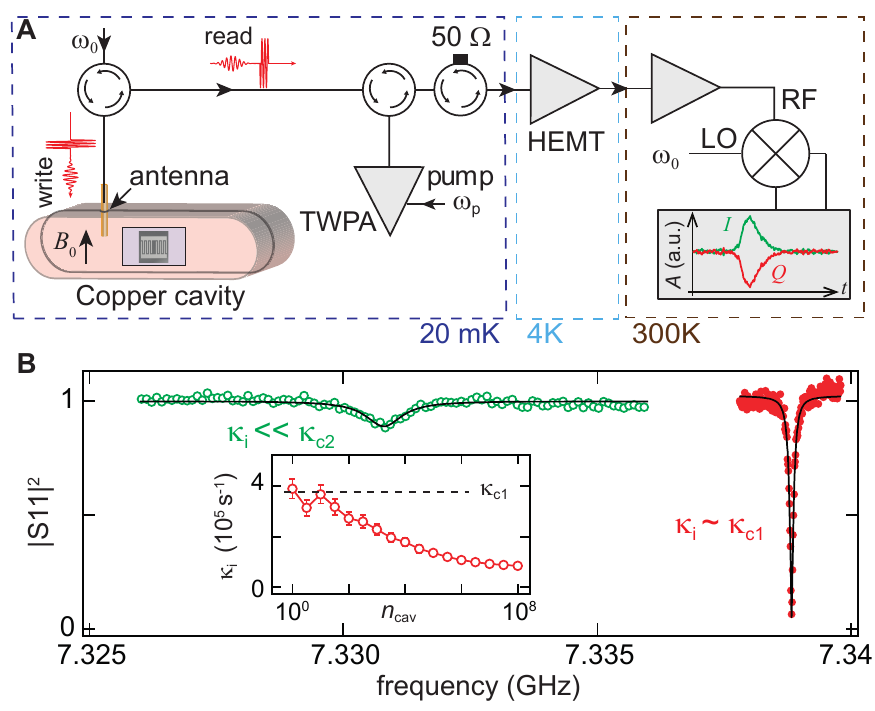}
  \caption{\label{fig:setup}
   \textbf{Measurement setup. (A)} The device is mounted inside a copper box to facilitate measurements in reflection via an antenna coupling. The reflected pulses and echo at the resonance frequency $\omega_0$ are first amplified by a TWPA followed by further amplification and homodyne detection at room temperature. \textbf{(B)} Main panel: Reflectance measurements of the resonance for two lengths of the antenna performed at an average intra-cavity photon number of one. Black curves are least square fits. Inset: Change in the resonator internal loss rate $\kappa_\text{i} $ as a function of average intra-cavity photon number $n_\text{cav}$. Measurements are done at $B_0 = 0$.
  }
\end{figure}

The spin-resonator hybrid device is measured at the base temperature of a dilution refrigerator in a setup as shown in Fig.~\ref{fig:setup}\textbf{(A)}. The substrate is a natural silicon wafer with an epitaxially grown 700~nm surface layer of isotopically purified 99.95\% $^{28}$Si. This was implanted with Bi ions at energies of 40, 80, 120, 200, and 360~keV with a total fluence of $1.1\times 10^{12}$~cm$^{-3}$ and annealed at $800^{\circ}$C for 20~min in an N$_2$ atmosphere (identical to the $^{28}$Si sample used in Ref.~\onlinecite{weis_electrical_2012}). The resonator was patterned by electron-beam-lithography followed by evaporation of a 50~nm aluminum thin-film and liftoff. Prior to fabrication, the substrate was cleaned with a piranha solution, and oxygen plasma ashing was used shortly before thin-film deposition. The resonator inductance is a $700~\mu$m long and $5~\mu$m wide wire, which is shunted by a co-planar capacitor consisting of 12 inter-digitated fingers, $50~\mu$m wide and $50~\mu$m apart.

\begin{table}[b!]
\caption{Resonator parameters}
\label{table1}
\begin{center}
  \begin{tabular}{ | c | c |  }
    \hline
    Inductor width/length& 5~$\mu$m/700~$\mu$m  \\ \hline
    $\omega_0/2\pi$ at $B_0$ = 27~mT  & 7.336 GHz \\ \hline
     Resonator impedance  & 40 $\Omega$ \\ \hline
    $\kappa_c$ & $4 \times 10^5~\text{s}^{-1}$ \\ \hline
     $\kappa_i~(n_\text{cav}=1, B_0 = 27~\text{mT})$ & $9 \times 10^5~\text{s}^{-1}$  \\ \hline
     $\kappa~(n_\text{cav}=1, B_0 = 27~\text{mT})$ & $13 \times 10^5~\text{s}^{-1}$ \\ \hline 

  \end{tabular}
\end{center}
\end{table}

The device is mounted inside a copper box with single antenna coupling to allow measurements in reflection. Reflected signals, before demodulation at room temperature, are first amplified by a TWPA at 20~mK and then by a HEMT at 4~K. The length of the antenna can be tuned to control the coupling and hence tune the resonance frequency, in this case by $\sim 10$~MHz, which allows us to target the clock transition. Measured resonance curves are shown in the main panel of Fig.~\ref{fig:setup}\textbf{(B)} for two values of coupling rate $\kappa_c$. 

The magnetic field of $27$\,mT needed to bias the bismuth donor spins at their CT is large compared to the $10$\,mT critical field of bulk aluminum, and can therefore only be applied parallel to the sample. In our experiment, alignment is achieved mechanically before cool-down, since the field is applied by a single-axis coil, and residual misalignments are unavoidable. As a result, significant phase noise and frequency hysteresis $(\sim 1~\text{MHz})$ are observed in the resonator response. Moreover, the resonator frequency is seen to vary by $\pm 1~$MHz and $\kappa$ change by $\pm 20\%$ from one cooldown to another. Resonator phase noise translates into an additional decoherence channel for the spins, because it results in an uncontrolled relative phase between the $\pi/2$ and $\pi$ pulses. For the $T_2$ measurements of Fig.~2, this can be circumvented by averaging the echo magnitude and not quadratures, as is commonly done in EPR spectroscopy. However, magnitude averaging is not possible for the memory protocol of Fig.~3 since phase coherence between the input and output is essential; as a result, the data of Fig.~3 could only be acquired at times when the resonator phase noise was not too strong. 

The resonator internal loss rate $\kappa_{i}$ is observed to decrease with average intra-cavity photon numbers as shown in the inset of Fig.~\ref{fig:setup}\textbf{(B)} for $B_0=0$. Power-dependent losses are a well-known signature for microwave losses caused by a bath of Two-Level Systems (TLSs). In the context of our work, these TLSs compete with bismuth spins in photon absorption, see further below. Near 27~mT and at low power $n_\text{cav} \sim 1$, internal loss rate $\kappa_i$ is $9 \times 10^5~\text{s}^{-1}$ while at high input power, $n_\text{cav} \sim 10^6$ , $\kappa_\text{i} \sim \kappa_\text{c} = 4\times 10^5 \text{s}^{-1}$.

%%###################### Theory and simulations ############
\section{Theory and simulations}
The expectation values of the intra-cavity field operator $\langle a  \rangle = \alpha $ can be described via standard input-output theory for $N$ independent spins (labeled by index $m$) coupled to a common resonator with coupling rate $g_m$,
\begin{align}
   \label{eq:CavField}
\frac{\partial \alpha}{\partial t} = \sqrt{\kappa_c} \beta - \frac{\kappa}{2} \alpha  -i \sum_{m=1}^{N} g_m S_{+}^{(m)}, 
\end{align}
where $S_{+}^{(m)} = \left[ \langle \hat{\sigma}_x^{(m)} \rangle  +i \langle \hat{\sigma}_y^{(m)} \rangle \right] /2$ is the expectation value of the $m$ spin raising operator, $\beta$ the input field, $\kappa_c$ and $\kappa$ being the external and total coupling rate of the resonator. The spin dynamics is described by Bloch's equation

\begin{align}
    \label{eq:Bloch}
\frac{\partial S_{+}^{(m)}}{\partial t} = -i \Delta_m S_{+}^{(m)} + 2i  g_m S_z^{(m)} \alpha - S_{+}^{(m)} / T_2. 
\end{align}
Here $\Delta_m = \omega_m - \omega_0$ is the spin-cavity detuning and $T_2$ the spin coherence time. 

Simple analytical expressions can be obtained in the steady-state in the cases of single $g_0$, large inhomogeneous linewidth $\Gamma \gg \kappa$, small drive bandwidth $t_0^{-1} \ll \kappa$ and long $T_2$ coherence times. The absorption of a weak field is obtained by setting $S_z^{(m)}=1$, and the retrieval of a weak echo by setting $S_z^{(m)}=-1$ for the single-$\pi$ pulse protocol. The following expressions are then obtained~\cite{afzelius_proposal_2013}

\begin{align}
	\label{eq:alpha}
\alpha^\text{sto} &= \frac{\sqrt{\kappa_c}}{\kappa} \frac{2}{(1+C)}  \beta, \\ 
\alpha^\text{ret} &=  -\frac{\sqrt{\kappa_c}}{\kappa} \frac{4C}{(1+C)(1-C)} \beta
\end{align}
where $C = 4g^2_0 N/\kappa \Gamma$ is the ensemble cooperativity. In the case of a single refocusing pulse, a divergence appears at $C=1$ because the retrieval happens when the spins are in the excited state and can thus behave as a transient maser, providing gain in the system. This reveals the limitation of the single refocusing pulse scheme. The full quantum memory protocol employs two $\pi$ pulses~\cite{afzelius_proposal_2013} such that spins are back to the ground state and in this case the retrieved field is $\alpha^\text{ret} =  - \frac{\sqrt{\kappa_c}}{\kappa} \frac{4C}{(1+C)^2} \beta$. For $\kappa \sim \kappa_c$ and $C=1$, impedance matching is achieved, implying that the absorption of the input field by the spins as well as a its retrieval after refocusing are complete. 

Disregarding any spin decoherence, and in the limit of $C \ll 1$, the echo amplitude is given by $\sqrt{\kappa_c} \alpha^\text{ret} \approx \zeta \beta$, with $\zeta = 4C (\kappa_c/\kappa)$, and the ratio of retrieved to input energy is provided by $\zeta^2$. This is equally true for a storage protocol including one or two $\pi$ pulses.

A complete time dependent numerical simulation includes characteristics of the spin ensemble such as the distributions of Larmor frequency $\rho_\delta(\delta)$ and coupling constant $\rho_g(g)$, normalized such that $\int dg \rho_g (g) = \int d\delta \rho_\delta (\delta) = 1$. We note that such characteristics directly affect both Rabi angles and Purcell relaxation rates and therefore the relative contribution to output signals for a given repetition rate. The Purcell relaxation time is given by $ T_1^{(m)} = \frac{\kappa}{4 g_m^2} \left [1 + \left(\frac{2\Delta_m}{\kappa}\right)^2 \right]$.  Because the ensemble inhomogeneous linewidth is much larger than the cavity-linewidth, we assume $\rho_\delta(\delta)$ to be a square function of size $\sim 10 \kappa$ centered around the resonator frequency.

%%###################### Reset by Purcell relaxation ############
\section{Purcell limited $T_1$}
\begin{figure}[tbph]
  \includegraphics[width=\columnwidth]{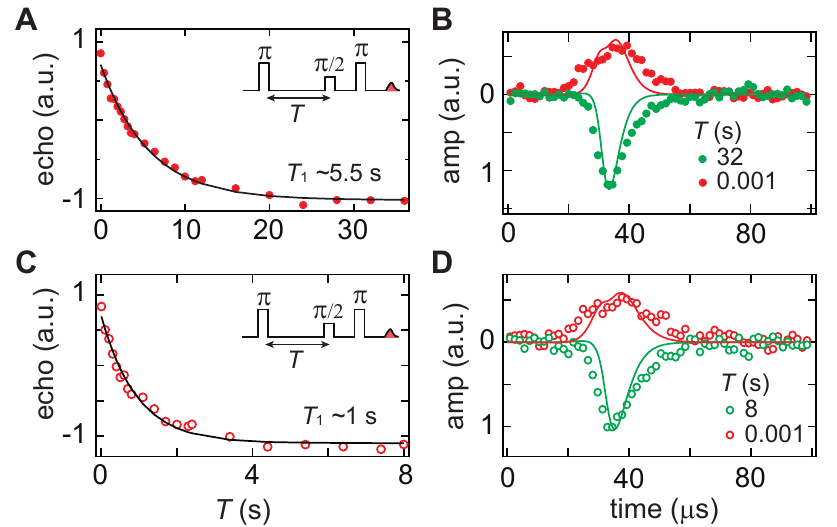}
  \caption{\label{fig:T1}
   \textbf{Purcell energy relaxation. (A,~C)} Inversion recovery measurements in symbols and fit as solid black curves. \textbf{(B,~D)} Measured echoes at different time delays for the clock transition ($B_0 = 27$~mT) and for the first transition ($B_0 = 1.4$~mT). The solid lines are numerical simulations for a fixed $g_0$. 
  }
\end{figure}

In order to reset the spin ensemble in its ground state in-between two experimental sequences, the repetition time is chosen to be longer than the spin energy relaxation time $T_1$. In our experimental conditions, spin relaxation is purely radiative, with $T_1 = \kappa/4g_0^2$ \cite{bienfait_controlling_2016}. Spin relaxation curves acquired using inversion recovery sequence with square pulses, each 2~\textmu s in duration, are shown in Fig.~\ref{fig:T1} for the first transition ($\ket{4,4}\leftrightarrow\ket{5,-5}$) and for the clock transition ($\ket{4,-1}\leftrightarrow\ket{5,0}$ and $\ket{4,0}\leftrightarrow\ket{5,-1}$). The $T_1$ extracted from exponential fits directly yield the single spin-photon coupling strength $g_0/2\pi = 76~(40) \pm 8(4)$~Hz where we have used the low power total decay rate $\kappa$ of $9~(13) \times 10^5~\text{s}^{-1}$ for the first transition (CT). The values extracted are consistent with transition matrix elements $ \langle 0 |S_x|1\rangle=0.48~(0.25)$ for the first transition (CT). Together with these parameters, we are able to predict the complex echo shapes without any adjustable parameter at different delays due to dependence of Rabi angle and $T_1$ on the detuning, see Fig.~\ref{fig:T1}\textbf{(B,D)}.

%%###################### Strain and spectrum ############
\section{Bismuth spin spectrum}
\begin{figure}[t!]
  \includegraphics[width=\columnwidth]{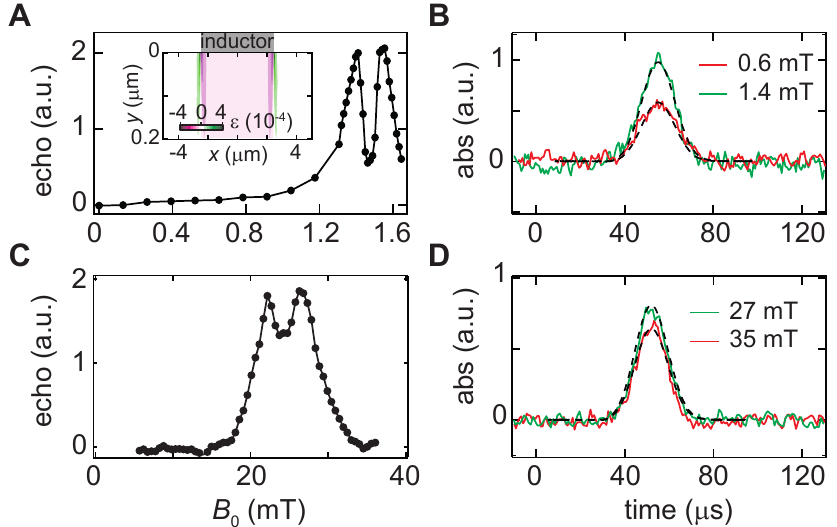}
  \caption{\label{fig:TLS}
  \textbf{Absorption from two-level systems. (A,~C)} Spectroscopy of bismuth spins near the first and the clock transition. Inset: A colormap of finite element simulation of hydrostatic component of strain. \textbf{(B,~D)} Absorbed intra-cavity fields at different magnetic fields. Measurements are done using gaussian pulses with roughly 240~photons in the input field. Dashed curves are calculated gaussians.  
  }
\end{figure}

The lack of in-situ frequency tuning prevents a complete spectrum measurement near the clock transition. The spin-spectrum near the first transition can however be obtained by sweeping the magnetic field  (see the main panel of Fig.~\ref{fig:TLS}\textbf{(A)}). The spectrum displays a split-peak shape with an inhomogeneous linewidth of $\sim 0.2~$mT or 5~MHz, much larger than the expected $\sim 1~$kHz dipolar linewidth and than the $100$\,kHz broadening due to the bath of residual $^{29}$Si nuclear spins~\cite{george_electron_2010}.

The split-peak spectrum was explained in Ref.~\onlinecite{pla_strain-induced_2018} by the dependence of bismuth donor hyperfine constant on the hydrostatic component of the strain $\epsilon$, with  $(1/2\pi)\text{d}A/\text{d}\epsilon = 29~$GHz. In our device, strain is due to the ten-times smaller thermal expansion coefficient of silicon relative to aluminum. A colormap of hydrostatic strain computed using COMSOL software is shown in the inset of Fig.~\ref{fig:TLS}\textbf{(A)}. Three colors of the colormap roughly denote three species of spins. Pink areas represent negatively strained spins just below the wire $|x|<2.5~\mu \text{m}$ and form the left peak and the left tail of the spectrum. White areas $|x|>2.5~\mu \text{m}$ are unstrained and form for the right peak. Green areas are positively strained spins making the tail of the right peak. Since the strain is magnetic field independent, we can identify the spin location for the measurements near CT. Given $\omega_0/2\pi$ is below the unstrained spin transition frequency by $ f_\Delta \sim -2.6~$MHz, spins must be located below the wire ($x = 0 \pm 0.5~\mu$m). This is supported by the black curve in Fig.~2\textbf{C} reproduced using the spectrum near first transition and measured $f_\Delta$ near CT.

%%###################### TLS ############
\section{TLS absorption}
Two level systems (TLSs) already show up in the power dependence of internal quality factor, see the inset of Fig.~\ref{fig:setup}. We extract their contribution quantitatively by measuring the absorbed field at magnetic fields where spins are resonant or non-resonant with the resonator frequency $\omega_0$. To get the absorbed field, we subtract the reflected field measured with a repetition time $t_\text{rep} = 3T_1$ from the one taken at $t_\text{rep} \ll T_1$, such that spins are either polarized or saturated, respectively. The measurements at different magnetic fields are plotted as solid lines in Fig.~\ref{fig:TLS}\textbf{(B,D)}. For the first transition (CT) we observe that only 40~(20)\% field absorption comes from spins, the rest being from the TLSs. The difference in absorption fraction between the first and the CT transition is consistent with the expected reduction in cooperativity. Indeed, at the CT, the matrix element is twice smaller than at the first transition, contributing to a reduction by a factor of $4$ since $C$ is proportional to $g_0^2$, whereas the number of spins is doubled because of the transition quasi degeneracy, leading to an overall cooperativity reduction by a factor of $2$, as observed. 
All relevant plots in the main text and the Supplementary materials have been corrected for the contribution from TLSs.

%%###################### storage and retrieval ############
\section{Number of spins}
A rough estimate of the spin number $N$ can be made using the bismuth implantation profile, the device geometry and the spin linewidth. Taking the nominal concentration of $8\times 10^{16}~\text{cm}^{-3}$, there are $2.8 \times 10^7$ spins in a box of volume $5~\mu\text{m} \times 0.1~\mu\text{m} \times 700~\mu\text{m} $ (volume only below the wire forming the left peak of the split spectrum in Fig.~\ref{fig:TLS}\textbf{A}). These spins are distributed among states of the lower-energy $F=4$ manifold with a linewidth of $\Gamma/2\pi \sim 2.5$~MHz, thus there are $\sim 1.2 \times 10^6$~spins/MHz per level of the $F=4$ manifold.

\begin{figure}[t!]
  \includegraphics[width=\columnwidth]{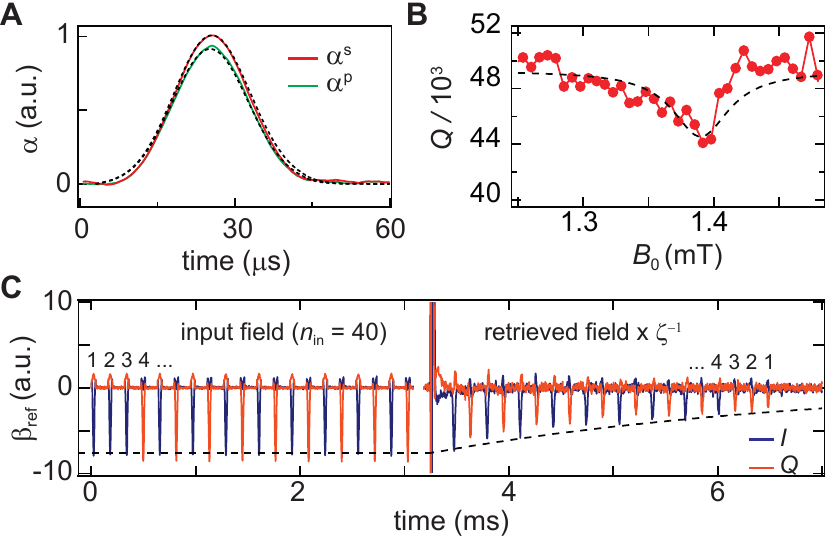}
  \caption{\label{fig:peak1L}
  \textbf{Storage and retrieval of microwave photons near the first transition. (A)} Measured intra-cavity field for the cases of spins saturated ($\alpha^s$, red) and polarized ($\alpha^p$, green). \textbf{(B)} Change in total quality factor of the resonator due to absorption from spins measured at average photon number of $n_\text{cav} \sim 0.1$.  \textbf{(C)} Measured input fields and retrieved field quadratures $I$ and $Q$. The retrieved field is multiplied by $\zeta^{-1}$, where $\zeta = 4C(\kappa_c/\kappa) = 0.14$ is the memory field efficiency at time $\ll T_2$. Dashed curves in all panels are numerical simulations. 
  }
\end{figure}

For a quantitative extraction of $N$, we measure the change in the intra-cavity field $\alpha^\text{sto}$ when spins are saturated or polarized. In our case, a pulse sequence can be simply repeated on a time scale much shorter than $T_1$ to saturate the spins. Experimentally measured intra-cavity fields are shown in  Fig.~\ref{fig:peak1L}(A)  for the first transition from which we deduce $C \approx 0.08$. This value is consistent with the numerical simulations shown as dashed lines and yielding the number of spins coupled to the resonator as,  $N/(\Gamma/2\pi) \approx 5 \times 10^5~\text{spins}/$MHz for $g_0/2\pi \approx$76~Hz. The same method has been employed in the main text (Fig.~\ref{fig:memory}\textbf{(A)}) to extract $ 10^6~\text{spins}/$MHz near the CT.

An independent approach to quantify spin absorption is to measure change in the resonator quality factor as a function of the magnetic field as shown in the Fig.~\ref{fig:peak1L}(A). The approach has the advantage that the TLSs do not react to magnetic fields. Changes in $Q=\omega_0/\delta \omega_B$ can be derived using input-output theory~\cite{schuster_high-cooperativity_2010}

\begin{equation}
    \label{eq:Qtot}
    \delta \omega_B = \kappa + \gens^2 \frac{\Gamma/2}{(\Delta_s^2 + \Gamma^2/4)},
\end{equation}
where $\kappa$ is the total decay rate of the resonator away from spin resonance,  $\Delta_s = \omega_0 - \omega_s$ the detuning between spins and the resonator,  $\Gamma$ the spin ensemble linewidth and $\gens = g_0 \sqrt{N}$. The change in total quality factor $Q~(=\omega_0/\delta \omega_B)$ as a function of $B_0$ near the first transition is plotted in Fig.~\ref{fig:peak1L}\textbf{(B)}. The dependence roughly follows the asymmetric shape seen in echo detected spectroscopy. From the Eq.~\ref{eq:Qtot}, we estimate $N = 10^6$ spins distributed in a Lorentzian spectrum of FWHM linewidth $\Gamma/2\pi = 1$~MHz. This estimate of $N/\Gamma$ is within a factor of 2 of the one extracted using the absorbed intra-cavity field.  

%%####################################################
\section{Field absorption and retrieval}
In addition to the data shown in the main text, we also demonstrate the storage and retrieval of weak coherent fields at the first Bi:Si transition ($B_0 = 1.4$~mT, $\dfdb = 25$~MHz/mT) as shown in~Fig.~\ref{fig:peak1L}\textbf{(C)}. A train of microwave fields each containing less than 40 photons is incident on the spin ensemble and retrieved using a refocusing $\pi$ pulse. We see again that the retrieved signal maintains the phase relation with respect to the input field. The storage time is however much smaller than at the CT, since the coherence time is $T_2 = 7.5$~ms. The retrieval field efficiency $\zeta = 4C(\kappa_c/\kappa) = 0.14$ here is however larger than the value measured at CT due to lower internal losses $(\kappa_i =  5 \times 10^5~\text{s}^{-1}, \kappa_c =  4 \times 10^5~\text{s}^{-1})$ and stronger spin-photon coupling strength $g_0/2\pi = 76~$Hz.

%% ################# Noise measurements #######################

\section{Noise measurements}
Here, we detail the noise measurements reported in Fig.~4. Considering $N$ spins excited during the refocusing $\pi$ pulse, the number of photons $n_\text{SE}$ emitted due to their spontaneous emission between time $t$ and $t+dt$ is $ N dt/T_1$ for $t, dt\ll T_1$. In the Purcell limit, this yield the number of photons emitted by spontaneous emission during the echo, 
\begin{equation}
    n_\text{SE} = N\frac{4g_0^2}{\kappa}dt \sim N\frac{4g_0^2}{\kappa \Gamma} =  C,
\end{equation}
since the echo duration is $\sim 1/\Gamma$. 

We can define the field annihilation and creation operators in the output mode in which the echo is emitted as $a = \int u(t - t_0) a_\text{out}(t) dt$ and $a^\dagger = \int u(t - t_0) a_\text{out}^\dagger (t) dt$, where $t_0$ is the mean echo arrival time, and $u(t)$ is the envelope function defining the mode, verifying $\int |u(t)|^2 dt = 1$. We take $u(t)$ as the Gaussian envelope of both the input pulse and the echo, i.e. $u(t) = \beta_\text{in} (t) / \sqrt{\int \beta_\text{in}^2(t)dt}$. With this definition, the quadrature $X_a = (a + a^\dagger)/2 $ verifies $\delta X_a ^2 = 1/4$ for a coherent state and for the vacuum, and $\delta X_a ^2 = 1/4 (1 + 2C)$ for a thermal state containing $C$ photons on average.

The total output noise also includes the amplifier noise. Amplification is modelled as transforming input mode $a$ into output mode $b = \sqrt{G} a + \sqrt{G-1} a_\text{id}^\dagger$, where $G$ is the power gain, and $a_\text{id}$ the idler operator. Defining the output mode quadrature as $X_b = (b + b^\dagger)/2$, we get $\langle b \rangle = \sqrt{G} \langle a \rangle$, and $\delta X_b^2 = G \delta X_a^2 + (G-1) \delta X_\text{id}^2$. 

We then write $\delta X_\text{id}^2 = \frac{1}{4}(1 + 2 \bar{n}_\text{id})$. In this equation, a non-zero value of $\bar{n}_\text{id}$ accounts for amplifier non-ideality, but also for losses in-between the sample and the amplifier as well as added noise by the following amplifiers of the detection chain. In the limit of large gain, the total noise referred to the amplifier input thus writes $\delta X^2 = [G/4 + (G-1)(1 + 2 \bar{n}_\text{id})/4]/G \simeq (\bar{n}_\text{id} + 1)/2$ for vacuum or a coherent state, and $\delta X^2 =  (\bar{n}_\text{id} + C + 1)/2$ for a thermal state of $C$ photons on average.

As explained in the main text, we find $\bar{n}_\text{id} = 3.5 \pm 1.7$, a reasonable value compared to those measured in similar circuit QED setups. This consistency check gives us confidence both in the determination of field amplitudes and photon numbers in absolute units, and confirms that the echo emission occurs indeed with negligible added noise.

\end{document}